\documentclass[prb,twocolumn,preprintnumbers,amsmath,amssymb,superscriptaddress]{revtex4}
\usepackage{graphicx}
\usepackage{latexsym}
\usepackage{amsmath}
\usepackage{graphics}
\usepackage{amssymb}
\usepackage{layout}
\usepackage{verbatim}
\usepackage{amsfonts,epsfig}
\newcommand{\ket}[1]{\left|#1\right>}
\newcommand{\bra}[1]{\left< #1 \right|}
\newcommand{\beq}{\begin{equation}}
\newcommand{\eeq}{\end{equation}}

\begin{document}

\title{Optical orientation of a single Mn spin in a quantum dot: Role of carrier spin relaxation}
\author{{\L}ukasz Cywi{\'n}ski} \email{lcyw@ifpan.edu.pl}
\affiliation{Institute of Physics, Polish Academy of Sciences, al.~Lotnik{\'o}w 32/46, PL 02-668 Warszawa, Poland}
\date{\today }

\begin{abstract}
In order to explain the recently observed phenomenon of optical orientation of a single Mn spin residing inside a CdTe quantum dot, a process of Mn spin relaxation with characteristic timescale of tens of nanoseconds had been invoked. We show that after taking into account the mixing of states of the exciton and the Mn spin (due to the sp-d exchange interaction), the observed Mn optical orientation time can be explained by invoking only the processes of carrier spin relaxation.  
\end{abstract}

\maketitle

\section{Introduction}
Quantum dots containing a single Mn impurity have been a subject of growing interest in  the last few years.\cite{Besombes_PRL04,Leger_PRL06,Kudelski_PRL07,Krebs_PRB09}
Optical orientation of a single Mn spin inside a CdTe quantum dot (QD) has  been observed recently. \cite{LeGall_PRL09,Goryca_PRL09,Besombes_SSC09,LeGall_PRB10} In these experiment circularly polarized light was creating a spin-polarized exciton (X) in the dot, and upon constant illumination the Mn spin became polarized on a timescale of $\tau_{\text{Mn}} \! < \! 100$ ns. One mode of excitation is quasi-resonant, through an excited state of the dot in Ref.~\onlinecite{LeGall_PRL09} or through an exciton transfer from a resonantly excited nearby Mn-free dot in Ref.~\onlinecite{Goryca_PRL09}. Alternatively, one of the states of the Mn+X complex can be resonantly driven, as in Ref.~\onlinecite{LeGall_PRB10}. This mode of operation was originally proposed in Ref.~\onlinecite{Govorov_PRB05}, and it is the focus of this paper. 
Specifically, we will consider the situation from Ref.~\onlinecite{LeGall_PRB10} in which the highest-energy line of the X+Mn complex is excited with $\sigma_{-}$ polarized light, thus creating a dominantly $\ket{-5/2;-1}$ state (written in the basis of the $S^{z}$ component of the Mn spin and the $J^{z}$ projection of the total spin of the exciton). Due to such an excitation the population of the $\ket{-5/2}$ Mn level was observed to decrease on the timescale of less than 100 ns.

These recent experimental achievements could pave the way to optical control of the Mn spin state (e.g.~being able to intialize the Mn spin in each of its six states, as proposed in Ref.~\onlinecite{Reiter_PRL09}). 
The physical mechanism of the optical orientation of the Mn spin remains, however, unclear, and its proper understanding will most probably be crucial for further experimental developments. The goal of this paper is to elucidate a possible microscopic mechanism of Mn optical orientation.
 
The ``intrinsic'' relaxation of Mn due to spin-lattice interaction (Mn spin flip due to scattering with phonons) is well known to be very slow for isolated Mn spins,\cite{Scalbert_pssb96,Dietl_PRL95} e.g.~spin-lattice relaxation times longer than a microsecond were observed in dilute samples at $T\! = \! 5$ K and at magnetic field of about $10$ T in Ref.~\onlinecite{Strutz_PRL92}. It should be stressed that the high-field relaxation times are relevant to the case of Mn interacting with a confined exciton, which splits the Mn spin levels via the sp-d exchange interaction. The relatively fast relaxation of isolated Mn spins observed recently at  zero $B$ field\cite{Goryca_relaxation_PRL09} is possibly relevant when the exciton is absent.

 On the other hand, the phonon-induced processes of carrier spin relaxation (spin flips of the electron, of the hole, or of the whole exciton) were predicted to be quite effective for large energy transfer involved in a spin-flip.\cite{Khaetskii_PRB01,Woods_PRB04,Tsitsishvili_PRB03,Tsitsishvili_PRB05,Roszak_PRB07} In the Mn-doped dot the spin splitting of the carrier states is enhanced by the sp-d exchange interaction, and clear signatures of both exciton and hole spin relaxation were observed there.\cite{LeGall_PRB10} 
It is therefore clear that the spin flips of the carriers occur on timescales relevant to the Mn optical pumping process. On the contrary, the existence of a fast (on a time-scale of tens of ns) process of ``intrinsic'' Mn spin flip in the presence of the exciton is still somewhat controversial. Such a process was included in the model used in Ref.~\onlinecite{LeGall_PRB10} in order to explain the $\tau_{\text{Mn}} \sim 70$ ns timescale of Mn orientation (it was also used in the original proposal\cite{Govorov_PRB05} of optically orienting the Mn spin by driving one of the six X+Mn transitions). As mentioned above, it is highly improbable that the spin-lattice interaction can account for such a fast process. Mn spin relaxation time of the order of 10 ns was observed in Ref.~\onlinecite{Besombes_PRB08} and explained there by assuming that the Mn is coupled to extended electronic states from the wetting layer. 
This mechanism requires the presence of free hot photocarriers outside of the dot (which would scatter on the Mn spin), which should not be the case for (quasi)resonant excitation of a single dot. 

The goal of this paper is to show that it is not necessary to include the ``intrinsic'' rate of Mn spin relaxation in the presence of an exciton, $\Gamma_{\text{Mn-X}}$, into the description of the optical pumping process. The Mn optical orientation can occur due to carrier spin relaxation (specifically the hole spin relaxation in the case of experiment from Ref.~\onlinecite{LeGall_PRB10}) and mixing of the exciton and Mn states due to sp-d exchange interaction. Because of the latter the eigenstates of the X+Mn system are superpositions of states with different exciton $J^{z}$ and Mn spin $S^{z}$. When a high-energy X+Mn state is excited, the carrier spin relaxation leads to transition to lower-energy states having different $S^{z}$ composition, and subsequent spontaneous recombination of these states leaves the Mn spin changed. In other words, in order to achieve Mn spin orientation, it is enough to consider the carrier spin relaxation in the strongly coupled system of the carriers and the Mn spin.

The paper is organized in the following way. In Section 2 we introduce the Hamiltonian of the system, including the sp-d exchange interaction and the electron-hole exchange. In Section 3 we briefly discuss the possibility of Mn spin optical orientation without any spin relaxation in the system. This optical pumping mechanism turns out to be inefficient, but its discussion highlights the significance of mixing of X and Mn states via the sp-d exchange interaction. An impatient reader can skip this Section and proceed directly to Section 4, where we include the carrier spin relaxation in the system dynamics and show that it could account for the recent observations.

%%%%%%%%%%%%
%% THE HAMILTONIAN
%%%%%%%%%%%%%
\section{The Hamiltonian of a single Manganese spin interacting with an exciton in a quantum dot}  \label{sec:H}
The Hamiltonian at zero magnetic field is $\hat{H} = \hat{H}_{sp-d} + H_{e-h} $. The first term is the sp-d interaction
\begin{eqnarray}
\hat{H}_{sp-d}&  = & -A_{e}( \hat{S}^{z}\hat{s}^{z} + \frac{1}{2}[\hat{S}^{+}\hat{s}^{-} + \hat{S}^{-}\hat{s}^{+}]) \nonumber \\
& &  + A_{h}( \hat{S}^{z}\hat{\kappa}^{z}/2 + \frac{1}{2}[\epsilon\hat{S}^{+}\hat{\kappa}^{-} + \epsilon^{*}\hat{S}^{-}\hat{\kappa}^{+} ]) \,\, ,
\end{eqnarray}
where $\hat{S}^{i}$ are the operators of the Mn spin ($S\! = \! 5/2$), $\hat{s}^{i}$ are the electron spin operators, and $\hat{\kappa}^{i}$ are the Pauli matrices operating in the two-dimensional subspace of dominantly heavy hole states (the Kramers doublet of the lowest-energy hole states confined in the dot). 
They appear after taking the matrix elements of the p-d interaction $A_{h}\mathbf{\hat{S}}\cdot\mathbf{\hat{j}}/3$ (with $\mathbf{\hat{j}}$ being the spin-$3/2$ operator) within the subspace of two mostly heavy-hole (hh) states being confined in the QD. The finite admixture of the light hole (lh) states in the relevant low-energy states (due to e.g.~anisotropic strain\cite{Besombes_JAP07,Leger_PRB07}) leads to $\epsilon \! \neq \! 0$ allowing for the flip-flop between the hole spin and the Mn spin.
$A_{e}$ and $A_{h}$ are the exchange interaction energies for the electron and the hole (with our sign convention they are both positive). 

The second term is  the electron-hole exchange interaction,\cite{Bayer_PRB02} which  is written as 
\begin{eqnarray}
\hat{H}_{e-h} & = & \frac{\delta_{0}}{2}( \ket{1}\bra{1} + \ket{-1}\bra{-1} - \ket{2}\bra{2} - \ket{-2}\bra{-2} ) \nonumber \\
& & \!\!\!\!\!\!\!\!\!\!\!\!\!\!\!\!\!\!\!\!\!\!\!\!\!\! + \frac{\delta_{1}}{2}( \ket{1}\bra{-1} + \ket{-1}\bra{1}) + \frac{\delta_{2}}{2}( \ket{2}\bra{-2} + \ket{-2}\bra{2}) \,\, ,
\end{eqnarray}
where we have used the basis of the total exciton angular momentum along the $z$ axis $\ket{J^{z}=s^{z}_{e} + j^{z}_{h}}$, and we have approximately identified the two mostly hh-like states with $j^{z}_{h} \! =\! \pm 3/2$ (thus neglecting the small corrections due to the hh-lh mixing).
$\delta_{0}$ is the isotropic exchange splitting of the bright and dark excitons, $\delta_{1}$ is the splitting of bright excitons present in dots with broken cylindrical symmetry, and $\delta_{2}$ is giving the splitting of dark excitons. The last two terms come from $b_{i} (J^{i}_{h})^{3}s^{i}_{e}$ terms in the e-h exchange Hamiltonian, which are present due to the cubic symmetry of the lattice, and as such are breaking the cylindrical symmetry of the exchange Hamiltonian, thus leading to mixing of states with different $J^{z}$.

\begin{figure}[t]
\includegraphics[width=0.99\linewidth]{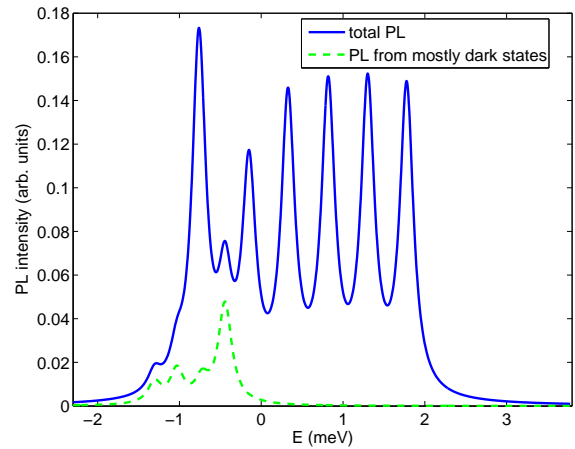}
\caption{(Color online) Photoluminescence spectrum of the dot with a single Mn spin calculated using the parameters given in the text. All levels are assumed to be equally populated. The contribution from the mostly dark states to the total PL is plotted with the dashed line. Line broadening of $0.1$ meV was used.}  \label{fig:PL}
\end{figure}

% VALUES OF THE PARAMETERS:
In the calculations below we will use the following parameters typical for small CdTe QDs. We take $A_{h} \! \equiv \! -\beta|\Psi_{h}(\mathbf{r}_{\text{Mn}} )|^2 \! = \! 0.8$ meV, with $\beta$ being the hole exchange integral, and $ \Psi_{h}(\mathbf{r}_{\text{Mn}})$ being the amplitude of the hole wavefunction at the Mn site. This value corresponds to $\approx \! 3$ meV width of sextuplet of bright exciton lines of X+Mn complex (see Fig.~\ref{fig:PL}). The value of $A_{e} \! \equiv \!  \alpha|\Psi_{e}(\mathbf{r}_{\text{Mn}})|^2 \! = \! 0.2$ meV follows from the ratio of $|\beta/\alpha| \! \approx \! 4$ in CdTe and from a somewhat arbitrary assumption of equal amplitude of electron ond hole wavefunctions at the Mn spin site (the hole is believed to be more weakly bound than the electron in CdTe dots, but its binding is enhanced in the presence of an electron,\cite{Besombes_JAP07} and it is unlcear which effect prevails). For the parameter giving the strength of the hole-Mn flip-flop we take a typical value of $|\epsilon| \! = \! 0.1$ deduced from the linear polarization of the QD photoluminescence\cite{Besombes_JAP07,Leger_PRB07} (the phase of $\epsilon$ determines the polarization axis, but it is irrelevant for the optical orientation effect discussed here).  For the electron-hole exchange energies we use\cite{Kazimierczuk_APPA09} $\delta_{0} \! =\! 1$ meV and $\delta_{1} = 0.1 $ meV, and we assume $\delta_{2} \! =\! 0.1$ meV (which probably is an overestimate). 

The energy spectrum of the above Hamiltonian is clearly visible in the photoluminescence (PL) signal from an excited dot.\cite{Besombes_PRL04,ROssier_PRB06} In the zeroth approximation, we can neglect $A_{e}$ (since it is usually much smaller than $A_{h}$), and also  put $\epsilon$, $\delta_{1}$, and $\delta_{2}$ equal to zero. Then the spectrum consists of 12 doubly-degenerate levels: 6 of them are bright (i.e.~they couple to light and contribute to the PL signal), and 6 are dark. Within each group the spacing of the levels is given by $A_{h}/2$, and the bright states are higher in energy by $\delta_{0}$ compared
 to the dark ones. In the calculation with the full Hamiltonian the main change is the ``brightening'' of the dark excitons, which occurs due to the flip-flop parts of the sp-d exchange interactions. This is shown in Fig.~\ref{fig:PL}, where 
 the total spectrum is deformed, and more than 6 peaks are  visible  (the additional ones corresponding to mostly dark excitons, the contribution of which to the total PL is shown by the dashed line). Nonzero values of $\delta_{1}$ and $\epsilon$ also lead to the linear polarization of the PL signal,\cite{Leger_PRB07,Besombes_JAP07} which is however irrelevant for this paper. 

%%%%%%%%%%%%
%% OPTICAL PUMPING WITHOUT SPIN RELAXATION
%%%%%%%%%%%%%
\section{Manganese optical orientation without spin relaxation} \label{sec:norelax}
First, let us note that Mn polarization can be induced by resonant optical pumping even in the absence of \emph{any} spin relaxation processes. The breaking of the cylindrical symmetry of the dot is only needed, i.e.~$\epsilon \! \neq \! 0$  and/or $\delta_{2} \! \neq \! 0$.  

We focus now on the excitation of the highest energy X+Mn state with $\sigma_{-}$ polarized light.\cite{LeGall_PRB10} The $\ket{e}$ state excited  in such a way contains a large amplitude of $\ket{-5/2;-1}$, but it also has admixtures of other states. For typical values of parameters the dominant admixtures are the ones of $\ket{-5/2; +1}$ state (caused by $\delta_{1}$ term mixing the bright excitons) and of $\ket{-3/2; -2}$ (caused by the electron-Mn flip-flop). In the 2nd order of perturbation theory the latter state contains an admixture of $\ket{-3/2;2}$ state due to $\delta_{2}$ interaction mixing the dark excitons, and in the third order the admixtures of $\ket{-1/2;1}$ and $\ket{-1/2;-1}$ states are created from $\ket{-3/2;2}$ state by electron and hole spin flip-flops with the Mn spin, respectively. The admixture of these $\ket{-1/2;\pm 1}$ states in the $\ket{e}$ state (with $b_{\pm 1}$ amplitudes) lead to a finite probability of the recombination of the $\ket{e}$ state into the empty dot $\ket{-1/2}$ state. Using the third order perturbation theory we have
\begin{eqnarray}
b_{1}  & = & \frac{ (A_{e}\sqrt{8}/2)\cdot (\delta_{2}/2) \cdot ( A_{e}\sqrt{5}/2 ) } {(2A_{e} + \frac{1}{2}A_{h} + \delta_{0} ) (\frac{1}{2}A_{e} + 2A_{h} + \delta_{0}) ( \frac{3}{2}A_{e}+\frac{3}{2}A_{h}) } \nonumber \\
b_{-1} & = & b_{1} \frac{3\epsilon}{2} \frac{A_{h}}{A_{e}} \,\, . \nonumber
\end{eqnarray}

In the simplest case, when $|b_{-1}| ~\! \ll \! |b_{1}|$ (corresponding to negligible hh-lh mixing), the re-absorption from the $\ket{-1/2}$ state can be neglected due to the optical selection rules ($\sigma_{-}$ light coupling only to $\ket{-1}$ excitons), and in the process of optical pumping of the $\ket{e}$ state the population of the $S^{z}$ levels is transferred from $\ket{-5/2}$ to $\ket{-1/2}$ by spontaneous emission of $\sigma_{+}$ polarized photons.

While the calculations of pumping dynamics with both $b_{\pm 1}$ amplitudes being finite show rich and interesting features (e.g.~the possibility of either depleting the $\ket{-5/2}$ level or increasing its occupation, depending on the values of $b_{\pm 1}$ and other parameters), one can quickly see that this kind of process is incapable of explaining the experimental timescale of Mn optical orientation. In the limit of $\epsilon\! = \! 0$ we can write rate equaitons for 3 levels (occupations of $\ket{e}$ state and the two empty dot states $\ket{-5/2}$ and $\ket{-1/2}$). With generation rate $G$ and spontaneous recombination rate $\Gamma_{\text{rec}}\! = \! 1/T_{\text{rec}}$, and with $\ket{e} = a\ket{-5/2;-1} + b\ket{-1/2;1} + ...$ (with other admixed states being optically inactive or having negligible amplitudes), we have the equations for occupation probabilities
\begin{eqnarray}
\frac{d p_{e}}{dt} & = & -G|a|^{2}p_{e} -\Gamma_{\text{rec}}p_{e} + G|a|^{2}p_{-5/2} \,\, , \\
\frac{d p_{-5/2}}{dt} & = & G|a|^{2}p_{e} +\Gamma_{\text{rec}}|a|^{2}p_{e} - G|a|^{2}p_{-5/2} \,\, , \\
\frac{d p_{-1/2}}{dt} & = &  \Gamma_{\text{rec}}|b|^{2}p_{e} \,\, .
\end{eqnarray}  
For strong driving, $G\! \gg \! \Gamma_{\text{rec}}$, we get for times longer than $1/G$ that $p_{-5/2} \! \approx \! \frac{1}{4}\exp(-\Gamma_{\text{rec}}|b|^2 t/2)$, which gives the Mn orientation timescale $\tau_{\text{Mn}} = 2T_{\text{rec}}/|b|^2$. With the parameters used here one gets $\tau_{\text{Mn}} \approx 10^{7} T_{\text{rec}}$, while in the experiment $\tau_{\text{Mn}} \! < \! 10^{3} T_{\text{rec}}$ was seen (using the value of $T_{\text{rec}} \! =\! 200$ ps).

%%%%%%%%%%%%%%%%%%%%%%%%%%%%%%%%%%%%%%%%%%%%%%%%
%% OPTICAL ORIENTATION VIA CARRIER SPIN RELAXATION
%%%%%%%%%%%%%%%%%%%%%%%%%%%%%%%%%%%%%%%%%%%%%%%%
\section{Manganese optical orientation with carrier spin relaxation}  \label{sec:relax}
The optical orientation mechanism described above is inefficient because it relies on very small $\delta_{2}$-induced mixing of the dark excitons, and also because both of the flip-flop related admixtures involve the energy denominators $\Delta E \! > \! \delta_{0}$, with the latter being much larger than the off-diagonal couplings $A_{e}$ and $\epsilon A_{h}$.
Much more efficient optical orientation can be obtained when we include the processes of carrier spin relaxation. A phonon-induced spin-flip of an electron (a hole)  leads to a transition from a  bright state $\ket{m; \pm 1}$ to a  dark state $\ket{m ;\pm 2} $ ($\ket{m;\mp 2}$). 
The mostly dark eigenstates of the full Hamiltonian contain admixtures of bright states with $m' \! = \! m\pm 1$ appearing in  the first order of the perturbation theory. The electron-Mn flip-flop terms 
are connecting the $\ket{m; \pm 2}$ state to $\ket{m\pm 1; \pm 1}$, while the hole-Mn flip-flop terms $\sim \! \epsilon A_{h} \hat{\kappa}^{\pm}\hat{S}^{\mp}$ are connecting it to $\ket{m\pm 1; \mp 1}$.

From the resonantly excited  state $\ket{e} \! \approx \! \ket{-5/2; -1}$ the electron spin relaxation leads to $\ket{-5/2; -2}$ state, which is is \emph{not} coupled by sp-d exchange to any other states, and in the first order of perturbation theory does not have any admixtures of states with flipped Mn spin.
We are thus led to consider the possibility  of the hole spin relaxation event, which leads to a transition into the state $\ket{r} \approx a\ket{-5/2; 2} + b_{e}\ket{-3/2; 1} + b_{h}\ket{-3/2,-1}$, with the amplitudes of other states being much smaller.  
The main admixture amplitudes are
\begin{eqnarray}
b_{e} \approx \frac{A_{e}\sqrt{5}/2}{\delta_{0}-2A_{e}+A_{h}/2} \,\, , \\
b_{h} \approx \frac{ \epsilon A_{h}\sqrt{5}/2}{ \delta_{0} + 2A_{h} - A_{e}/2 } \,\, .
\end{eqnarray}
For typical parameters we get $b_{e} \! > \! b_{h}$ (e.g.~with values used here we have $b_{e} \approx 0.2$, $b_{h}\approx 0.04$). 

The presence of hole spin relaxation was seen in Ref.~\onlinecite{LeGall_PRB10}, where it was shown that excitation of $\ket{1/2; +1}$ state was leading to the strongest PL from the ``dark'' state $\ket{1/2; -2}$, which was being populated by hole spin relaxation from the initial state. The optical activity of the mostly dark states is also visible in the calculated PL spectrum shown in Fig.~\ref{fig:PL}, where the PL signal from states having mostly dark character is plotted with the dashed line. With the parameters employed here, the dark states most strongly mixed with the bright states are the ones with dominant $\ket{-1/2; \pm 2}$ character (with energy $\approx \! -0.44$ meV, see the strongest ``dark'' transition in Fig.~\ref{fig:PL}), which contain large admixtures of $\ket{-3/2; \mp 1}$ states caused by the hole-Mn flip-flop term allowed by hh-lh mixing.

The rate equations for the populations of $\ket{e}$, $\ket{r}$, $\ket{-5/2}$, and $\ket{-3/2}$ levels (the $\ket{-1/2}$ level considered previously is neglected here, since its pumping has been shown in the previous Section to occur on a much longer timescale) are
\begin{eqnarray}
\frac{d p_{e}}{dt} & = & -(G  + \Gamma_{\text{rec}} +\Gamma_{\text{h}}) p_{e}  + Gp_{-5/2} + \Gamma^{'}_{\text{h}}p_{r}\,\, , \\
\frac{d p_{-5/2}}{dt} & = & ( G +\Gamma_{\text{rec}}) p_{e} - Gp_{-5/2} \,\, , \\
\frac{d p_{-3/2}}{dt} & = &  \Gamma_{\text{d}}p_{r} \,\, ,\\
% \,\,\,, \,\,\,\,  
\frac{d p_{r}}{dt} & = &   \Gamma_{\text{h}}p_{e}   -(\Gamma_{\text{d}} + \Gamma^{'}_{\text{h}}) p_{r} 
\end{eqnarray}
where the spontaneous recombination rate of the ``dark'' exciton is $\Gamma_{\text{d}} \approx (|b_{e}|^{2} + |b_{h}|^{2})\Gamma_{\text{rec}}$, $\Gamma_{\text{h}}$ is the hole relaxation rate, and $\Gamma^{'}_{\text{h}} \! =\! \exp(-\Delta E/k_{B}T)\Gamma_{\text{h}}$ is the rate for the  hole spin flip from the dark state back to the bright state.  $\Delta E \! \approx\!  \delta_{0} + \frac{5}{2}A_{h}$ is the energy difference between the two states. At $T\! = \! 5$ K and for $\Delta E \! = \! 3 $ meV obtained from the parameters used here $\Gamma^{'}_{\text{h}} \! =\! 10^{-3} \Gamma_{\text{h}}$. However, even with  larger $\Gamma^{'}_{\text{h}}$ the results discussed below are changed very little, and we will put $\Gamma^{'}_{\text{h}} \! = \! 0$ from here on.

We start with the initial conditions of $p_{-5/2}(0) \! =\! p_{-3/2}(0) \! = \! 1/2$ and all the other $p_{i}(0) \! =\! 0$.
In the strong driving (saturation) regime ($G \! \gg \! \Gamma_{\text{rec}}$, $\Gamma_{\text{h}}$)  
we get that at times $t \! \gg \! G^{-1}$ we have  $p_{-5/2}(t) \! \approx \! \frac{1}{4}\exp(-\Gamma_{\text{h}}t/2)$, i.e.~the $\ket{-5/2}$ state get emptied on timescale of hole spin relaxation. Its population is shifted to $\ket{-3/2}$ and $\ket{r}$ levels. For the population of the former state we have
\beq
p_{-3/2}(t) \approx 1 + \frac{\Gamma_{\text{h}}e^{-\Gamma_{\text{d}}t} - \Gamma_{\text{d}}e^{-\Gamma_{\text{h}}t/2} }{2(\Gamma_{\text{d}} - \Gamma_{\text{h}}/2)} \,\, 
\eeq
when $\Gamma_{\text{d}} \! \neq \! \Gamma_{\text{h}}/2$, and $p_{-3/2}(t) \approx 1 - \frac{1}{2} e^{-\Gamma_{\text{d}}t}(\Gamma_{\text{d}}t+1)$ when $\Gamma_{\text{d}} \! = \! \Gamma_{\text{h}}/2$. 
Before we reach the times $t \! \gg \! 2\Gamma_{\text{h}}^{-1}$, $\Gamma_{\text{d}}^{-1}$ most of the initial population of $\ket{-5/2}$ will have moved to $\ket{-3/2}$ state. The driven transition becomes then optically inactive, and the optical orientation process is complete.

\begin{figure}[t]
\includegraphics[width=0.99\linewidth]{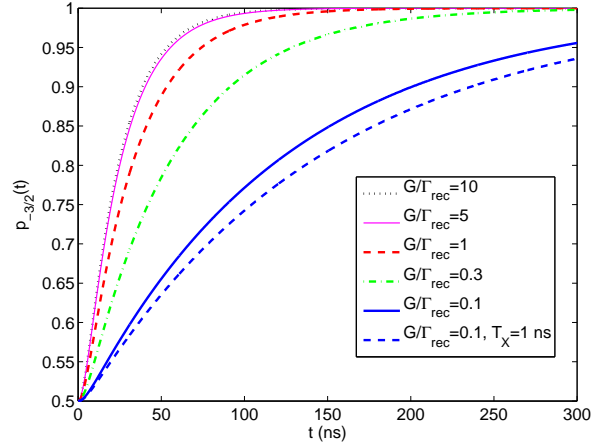}  
\caption{(Color online) Time dependence of the occupation of the empty dot $\ket{-3/2}$ Mn level upon pumping of the $\ket{-5/2,-1}$ transition with different light intensities (corresponding to different exciton generation rates $G$), with normalization $p_{-3/2}+p_{-5/2} \! =\! 1$. The spontaneous recombination rate is $\Gamma_{\text{rec}} \! = \! 5$ ns$^{-1}$ and the hole spin relaxation rate is $\Gamma_{\text{h}} \! = \! 0.1$ ns$^{-1}$. The exciton spin relaxation rate $\Gamma_{\text{X}}$ is assumed to be zero with exception of the dashed line for $G\! = \! 0.1\Gamma_{\text{rec}}$, for which $\Gamma_{\text{X}} \! = \! 1$ ns$^{-1}$. The other parameters are given in the text. 
}  \label{fig:G}
\end{figure}

\begin{figure}[t]
\includegraphics[width=0.99\linewidth]{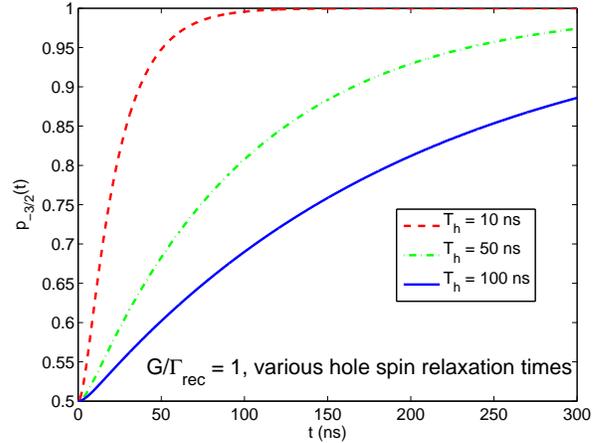}
\caption{(Color online) The same as in Fig.~\ref{fig:G}, only for $G/\Gamma_{\text{rec}} \!= \! 1$ and with various holes spin relaxation times $T_{\text{h}} \! = \! \Gamma_{\text{h}}^{-1}$. }  \label{fig:Tsr}
\end{figure}

With  $b_{e} \! \approx \! 0.2$ we get  the ``dark'' state recombination time $\Gamma_{\text{d}}^{-1} = 5$ ns, which is close to the observed value of $8$ ns.\cite{Besombes_PRB08} 
The calculations of $p_{-3/2}(t)$ for hole spin relaxation time $T_{\text{h}}\! =\! \Gamma_{\text{h}}^{-1} \! =\! 10$ ns are shown in Fig.~\ref{fig:G} for different exciton generation rates $G$. For $G\! \gg \! \Gamma_{\text{rec}}$ the analytical formulas given above are accurate, while at lower $G$ the rate equations have to be solved numerically. In Fig.~\ref{fig:Tsr} we show $p_{-3/2}(t)$ for various $T_{\text{h}}$ when $G\! = \! \Gamma_{\text{rec}}$.

The exciton spin relaxation\cite{Tsitsishvili_PRB03} leads to transitions from $\ket{e}$ to $\ket{f} \! \equiv \! \ket{-5/2,1}$ state with rate $\Gamma_{\text{X}} \! \equiv \! T^{-1}_{X}$. Including this effect in our rate equations is straighforward. However, as long as an assumption of $\Gamma_{\text{X}} \! \ll \! \Gamma_{\text{rec}}$ is made, the processes of exciton spin relaxation and subseqent spontaneous emission of $\sigma_{+}$ photon have very little impact on the optical orientation of the Mn spin. This is shown in Fig.~\ref{fig:G}, where a result for $G\! = \! 0.1 \Gamma_{\text{rec}}$ is shown also for $T_{\text{X}} \! = \! 1$ ns, and one can see that this leads to an insignificant slowing down of the orientation process. At higher $G$ the influence of finite $T_{\text{X}} \! > \! T_{\text{rec}}$ is even smaller.

The third process of carrier spin relaxation is the electron spin relaxation, which  leads to a transition into a dark state $\ket{d}\! = \! \ket{-5/2; -2}$ with recombination time of at least a couple hundreds of ns (using our rather large value of $\delta_{2}$, which is needed to bring about the mixing of this state with a bright one), which basically means that on the timescale of $\sim \! 100$ ns this state is perfectly trapping. If the electron spin relaxation time was faster than the hole spin relaxation time, then instead of the pumping of $\ket{-3/2}$ level the system would get trapped in the dark state $\ket{d}$. In the strong driving regime the transition corresponding to $\ket{-5/2; -1}$ state would become inactive on timescale of electron spin relaxation time $T_{\text{e}} \! = \! \Gamma_{\text{e}}^{-1}$, and instead of achieving optical orientation of the Mn spin one would obtain a dot with a very long-lived dark exciton trapped in it. The fact that this does not happen in Ref.~\onlinecite{LeGall_PRB10}, where the observations are consistent with the transfer of population between the Mn spin states, and not with the creation of stable dark exciton, shows that the electron spin relaxation is slower than hole and exciton spin relaxation processes.

%%%%%%%%%%%%
%% CONCLUSIONS:
%%%%%%%%%%%%%
\section{Conclusions}
We have shown that the experimental result from Ref.~\onlinecite{LeGall_PRB10}, the optical orientation of the Mn spin in tens of nanoseconds under a resonant driving of the highest-energy line of exciton+Mn complex, can be explained by the process of hole spin relaxation occurring on this timescale (which also has been observed in Ref.~\onlinecite{LeGall_PRB10}). The optical orientation occurs because the hole relaxation leads to a transition to an eigenstate of mostly dark character, which is  mixed with optically active states via sp-d exchange interaction. Since this admixture consists of states with a flipped Mn spin, the emission from the ``dark'' state populated by hole relaxation leads to a change of the spin polarization of the Mn ion. Consequently, the intrinsic processes of Mn spin relaxation (due to interaction with carriers in the wetting layer or phonons), do not have to be invoked in order to explain the optical orientation (this of course does not rule out their existence in some cases\cite{Besombes_PRB08}). 

Our analysis has also shown that the heavy-light hole mixing, while visibly present in the PL spectra, is not necessary to explain the Mn optical orientation (at least in the case of exciting the highest energy state of X+Mn complex). The three processes of hole, exciton, and electron spin relaxation, together with the electron-Mn exchange, can lead to quite a complicated behavior, with the excitation pattern considered here leading to a relatively simple dynamics. Further experiments involving resonant excitation of various lines of X+Mn complex, and observation of  PL signals induced in this way (as in Ref.~\onlinecite{LeGall_PRB10}), coupled with a calculation of X+Mn state mixing, might give more quantitative information on all the involved relaxation times. This knowledge could be then used in modeling of the situation from Ref.~\onlinecite{Goryca_PRL09}, where simultaneous excitation of many X+Mn levels leads to more complicated dynamics. 

One feature of the experimental results from Ref.~\onlinecite{LeGall_PRB10} which cannot be explained by the model proposed here is the saturation of the depletion of $\ket{-5/2}$ state at $75$ \%. Addressing this question is left for future research.

\emph{Note added.} The ``brightening'' of dark excitons due to the sp-d exchange interaction was very recently observed in Ref.~\onlinecite{Goryca_brightening}, where it was also shown that recombination from these states is an efficient channel of the Mn spin orientation.

\section{Acknowledgements}
The author is grateful to  T.~Dietl for discussions, reading of the manuscript, and commenting on it. Discussions with  {\L}.~K{\l}opotowski, M.~Goryca, O.~Krebs, and C.~Le Gall are also acknowledged. The financial support from the Homing programme of the Foundation for Polish Science supported by the EEA Financial Mechanism, from the EU FunDMS Advanced Grant of the European Research Council within the ``Ideas'' 7th Framework Programme, and from the European Union within European Regional Development Fund through grant Innovative Economy (POIG.01.03.01-00-159/08, ``InTechFun''), is gratefully acknowledged.

\end{document}